# Wireless Sensor Networks Attacks and Solutions

Naser Alajmi

Computer Science and Engineering Department, University of Bridgeport
Bridgeport, CT 06604, USA
nalajmi@my.bridgeport.edu

**Abstract—A few years ago, wireless sensor networks (WSNs) used by only military. Now, we have seen many of organizations use WSNs for some purposes such as weather, pollution, traffic control, and healthcare. Security is becoming on these days a major concern for wireless sensor network. In this paper I focus on the security types of attacks and their detection. This paper anatomizes the security requirements and security attacks in wireless sensor networks. Also, indicate to the benchmarks for the security in WSNs**

## I. INTRODUCTION

The security of wireless sensor networks is the area that has been discussed extremely through a few years ago. Networks have different applications. These applications comprise several levels of monitoring, tracking, and controlling. Wireless sensor networks consist of enormous number of small nodes. These nodes are deployed in some important areas. There are a group of applications that used for some purposes. So, in military application, sensor nodes include monitoring, battlefield surveillance and object tracking. The medical application, sensors can be helpful in patient diagnosis and monitoring. Most of these applications are deployed to monitor an area and to have a reaction when they record a sensitive factor [7]. Wireless sensor networks are emerging as both central new stage in the IT ecosystem and a rich area of active research involving hardware and system design networking, distributed algorithms, programming models, data management, security and social factors [2], [3]. Wireless sensor networks are vastly used in the area that is going to check for a particular task. Sensor nodes are liable to physical capture. Because the target of sensor nodes is low cost, tamper-resistant hardware is unlikely to take over.

The main goal of this paper is to find an energy efficient security solution thus to keep WSNs secured from any type of attacks. Also, this paper suggests several resolutions for the wireless sensor networks. Section II is analysis security requirements. Section III explains different security threats. Section IV makes a picture of benchmarks for WSN. Lastly, conclude on future works.

## II. SECURITY REQUIREMENTS

Security in wireless sensor networks has to be comprehensives a fundamental of requirements. These requirements are not only guarantee safeguard of sensitive data but also to achieve bounded resources in each sensor node, which remains the sensor network alive. Attacker motivation and vulnerabilities, and opportunities are two factors, which give the attacker possibility impact to the wireless sensor networks.

### A. Data Confidentiality

Data confidentiality is preserving the information hideaway from adversaries. The great way to keep the data invisible is to encrypt data with a secret key [1], [4]. The authorized can import data.

### B. Data Authentication

The basis for several applications in wireless sensor network is message authentication. Data authentication blocks any part that illegal from engage in the network alongside original nodes must be eligible to reveal from unauthorized nodes [1]. Also, data is important to make sure that started from the accurate source and the node that is claimed must be the end of a connection.

### C. Data Integrity

The adversaries have tried to change or modify the data. Therefore, data integrity makes sure the recipient who received message has not modified by unauthorized through transmission.

### D. Data Freshness

Data freshness means that the data is a fresh. So it guarantees that no old data or messages have been replayed. Although data integrity and confidentiality are assured, there is a need to make sure the freshness for each message. Furthermore, the malicious node does not reply or resend old or previous data.

### E. Access Control

Access control prevents unauthorized access to a resource. It should be able to prevent any participate in the network that is unauthorized.

| Secure Data Aggregation Identity Certificates | **Attacks:** 1. Denial of Service (DoS) | **Upper Layers:** Applicati on Layer |
|---|---|---|
| Secure Routing – Authorization, Monitoring | 2. Sybil 3. Blackhole Attack | **Middlew are Layers:** Transport & Network Layer |
| Shared Keys Protected Grouping | 4. Wormhole Attack 5. Flooding | **Lower Layers:** 1.Data Link |
| Encryption Processes | 6. Privacy 7. Intrusion Detection | 2.Physica l Layer 3.Hardwa re |

Figure 1.    Figure 1: Security at various levels in Sensor Network

## III.    SECURITY ATTACKS ON WSN

Wireless sensor networks are very weak and susceptible to many types of security attacks cause to the broadcast. Also, the other reason is put the sensor nodes in a dangerous environment whether in public area or battlefield. The security threats and attacks in wireless sensor networks as follows:

### A.    Sybil Attack:

Wireless sensor network is vulnerable to the Sybil attack. In such a case, a node can be more than one node using different identities of legal nodes. Therefore, a single node presents multi identities to other nodes in the network [6], (Figure 2). Sybil attack tries to degrade the integrity of data, security and resource utilization that the distributed algorithm attempts to achieve [2]. Authentication and encryption mechanisms can prevent an outsider to launch a Sybil attack on the wireless sensor network. Public key cryptography can avoid such an insider attack, but it is too costly to be used in the resource constructed sensor networks [4]. Identities must be verified so Karlof and Wagner [5] said that, might be done using public key cryptography, but generating and verifying digital signatures is beyond the capabilities of sensor nodes. Newsome, Shi, Song, and Perrig [13] indicated to several defenses against Sybil attack by using radio resource testing, verification of key sets for random key predistribution, registration and position verification in sensor network. The probability of Sybil node being detect is:

$$Pr(detection) = 1 - Pr(nondetection)^r_{1round}$$
$$= 1 - (1 - Pr(detection)_{1round})^r$$
$$= 1 - \left(1 - \sum_{all S,M,G} \frac{\binom{s}{S}\binom{m}{M}\binom{g}{G}}{\binom{n}{c}} \frac{S-(m-M)}{c}\right)^r$$

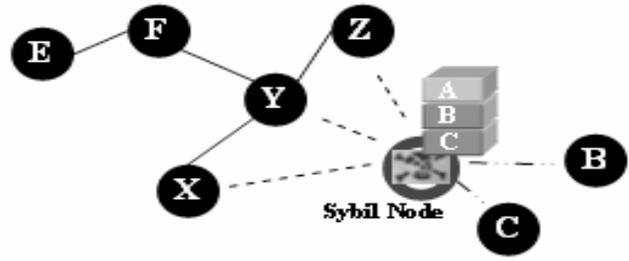

Figure 2.    Figure 2: Sybil Attack

### B.    Wormhole Attack:

Wormhole attack is a significant attack in which the attacker records the packet at single location in the network and tunnels those to another location. The transmitting of bits could be done selectively. Wormhole attack is an important threat to wireless sensor network, because this type of attack does not need compromising a sensor in the network or other. It could be implemented at the initial phase when the sensor launch to discover the information. The wormhole attack is showed in Figure (3). Two malicious nodes X1 and X2, connected by a powerful connection, create a wormhole. Node A and node B select the shortest route provided by the wormhole for send data. Data will be caught by the malicious nodes and then by the attacker [7].

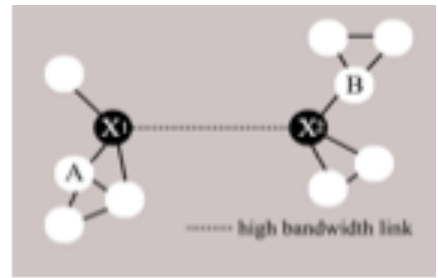

Figure 3.    Figure 3: Wormhole Attack

### C.    Denial of Service Attack:

Denial of service attacks can disrupt wireless transmission and occur either unintentionally in the form of interference, noise or collision at the receiver side or at the context of attacks [8]. There are some targets that attackers need to reach them such as network access, network infrastructure, and server application. DoS attack attempt to drain the resources available to the victim node by transferring extra needless data. Therefore, prevents users to accessing services. Denial of Service attack is meant not only for the adversary's seek to subvert, disrupt, or destroy a network but also for any event that diminishes a network's capability to provide a service [10]. Denial of Service attacks is created in different layers. In the physical layer the DoS attack could be jamming and tampering, at link layer, collision, exhaustion, unfairness, at network layer, neglect and greed, homing, misdirection, black hole and at transport layer DoS attack could be executed via malicious flooding and desynchronization [9].

The technique to prevent DoS attacks includes payment for network resources, pushback, strong authentication and identification of traffic [10]. There are some techniques to secure the reprogramming process thus one of them uses authentication flows. The choice for Denial of Service is the rekeying request packet. Hence it comes from the node only when any two consecutive keys are invalidated or lifetime of the keys has been expired. Therefore, if the rate of rekeying requests is coming frequently, then base station can conclude for possible DoS attack and drop the packet from the node for a configurable period of time [12]. The attacker find the re-keying request packet is a chance to send it again and make the DoS begins.

### D. HELLO Flooding Attack:

This type of attack uses HELLO packet as a weapon to encourage the sensor in networks. The attacker with a high radio transmission range and processing power sends HELLO packet to a number of sensor nodes thus they are separated in a large area within a wireless sensor network [5]. The sensors are thus impacted that the adversary are their neighbor. There are several agreements require HELLO packet radio node to node nearby to its own broadcasting. Attacker with power to begin track broadcast, so that the network each node is believed to attack its neighbors [11]. A malicious node with a powerful connection, which transfer HELLO messages to nodes that the malicious node is a neighbor and will send data to it [5], (Figure 3).

### E. Sinkhole Attack:

The sinkhole attack is an especially dangerous attack that prohibits the base station to gain entire and correct sensing data, consequently making a severe threat to the higher layer application (Figure 4). In a sinkhole, the adversary's goal is to lure nearly all the traffic from a particular area through a compromised node, creating a metaphorical sinkhole with the adversary at the center [5]. A compromised sensor node attempts to impact the information to it from any neighboring node. Thus, sensor node eavesdrops on each information is being communicated with its neighboring sensor nodes. Sinkhole attack works by making a compromised node look especially attractive to surrounding nodes with respect to the routing algorithm. For example, an adversary could spoof or reply an advertisement for an extremely high quality rout to the base station [5].

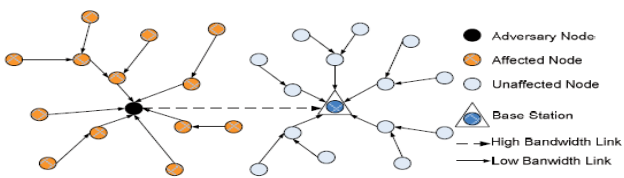

Figure 4.  Figure 4: Sinkhole Attack

## IV. SECURITY BENCHMARKS FOR WSN

In recent years, wireless sensor networks have grown and used for broad range of many applications such as weather, military aim tracking, and patient monitoring. Therefore, these sensor networks need protection from illegal attackers. There are some security benchmarks that sensor networks should be had.

### A. Encryption

In fact, most of wireless sensor network hold in an open area or dangers location, thus it susceptible to the network attacks. Eavesdropping or add messages into the network are significant to WSN [14]. It has to take over key methods that protect WSN such as message authentication codes, symmetric key encryption and public key cryptography [1].

### B. Data Partitioning

The technique of partitioning is to separate the data in networks into some or several parts. In wireless sensor networks, Deng J. [15] gives a solution to make sure the attacker cannot catch the information by using the data partitioning. Divide the data into multi packets so each packet transfers on a different route to nodes. At this point, the attacker tries to get all packets of a data from the network, thus it has to be capable to the entire networks. It is a perfect solution, but the energy consumption increased more than normal [7].

### C. Secure Data Aggregation

Transmit data in wireless sensor network increased than before. As a result, the most issue in network is data traffic. So, the cost is rising. To reduce the high cost and network traffic, wireless sensor node aggregates measurements before transferring to the base station [1]. Wireless sensor network architecture, aggregation carries out in many locations in the network. Aggregation locations should be secured [16].

### D. Cryptography

Symmetric key cryptography is a key that used in cryptography solutions in wireless sensor networks. Symmetric key is suitable and rapid to implement [7]. A cryptography method is used to prohibit some of the security attacks.

### E. Shared Keys

A better deal of the concentration in wireless sensor networks is the field of key management. WSN is a single in this feature because size, mobility and power constraints, [1]. There are four types of keys management, global key, pair wise key node, pair wise key group, and individual key. Each one of these keys is a solution to prevent attacks in wireless sensor networks.

## V. CONCLUSION

Security issue in wireless Sensor Network is more important than other issue. In recent years, security in WSN has frequently concerns. Wireless sensor networks are growing used in environment, commercial, health and military applications. This paper briefs a sort of   requirements that wireless sensor network have to be include and also introduce some of the security attacks. In addition wireless sensor network benchmarks.

## AUTHORS PROFILE

Naser Alajmi is a PhD student with the Computer Science and Engineering Department, University of Bridgeport, Bridgeport, CT, USA (e-mail: nalajmi@my.bridgeport.edu)